\documentclass[aps,pra,twocolumn,superscriptaddress,longbibliography]{revtex4-2}

\usepackage{amsmath}
\usepackage{bm}
\usepackage{amsfonts}
\usepackage{mathtools}
\usepackage{graphicx}
\usepackage{braket}
\usepackage{listings}
\usepackage{algorithm}
\usepackage{algorithmic}
\usepackage{color,ulem}
\usepackage{hyperref}
\usepackage{CJK}
\begin{document}
\begin{CJK*}{UTF8}{gbsn}

\title{Geometric quantum adiabatic methods for quantum chemistry}

\author{Hongye Yu (余泓烨)}

\affiliation{C. N. Yang Institute for Theoretical Physics, State University of New York at Stony Brook, Stony Brook, New York 11794-3840, USA}
\affiliation{Department of Physics and Astronomy, State University of New York at
Stony Brook, Stony Brook, New York 11794-3800, USA}

\author{Deyu Lu}
\email{dlu@bnl.gov}
\affiliation{Center for Functional Nanomaterials, Brookhaven National Laboratory, Upton, New York 11973, USA}
\author{Qin Wu} 
\affiliation{Center for Functional Nanomaterials, Brookhaven National Laboratory, Upton, New York 11973, USA}
\author{Tzu-Chieh Wei}
\email{tzu-chieh.wei@stonybrook.edu}
\affiliation{C. N. Yang Institute for Theoretical Physics, State University of New York at Stony Brook, Stony Brook, New York 11794-3840, USA}
\affiliation{Department of Physics and Astronomy, State University of New York at Stony Brook, Stony Brook, New York 11794-3800, USA}

\begin{abstract}
Existing quantum algorithms for quantum chemistry work well near the equilibrium geometry of molecules, but the results can become unstable when the chemical bonds are broken at large atomic distances. For any adiabatic approach, this usually leads to serious problems, such as level crossing and/or energy gap closing along the adiabatic evolution path. In this work, we propose a quantum algorithm based on adiabatic evolution to obtain molecular eigenstates and eigenenergies in quantum chemistry, which exploits a smooth geometric deformation by changing bond lengths and bond angles.  Even with a simple uniform stretching of chemical bonds, this algorithm performs more stably and achieves better accuracy than our previous adiabatic method [Phys. Rev. Research 3, 013104 (2021)]. It solves the  problems related to energy gap closing and level crossing along the adiabatic evolution path  at large atomic distances. We demonstrate its utility in several examples,  including H${}_2$O, CH${}_2$, and a chemical reaction of H${}_2$+D${}_2\rightarrow$ 2HD. Furthermore, our fidelity analysis demonstrates that even with finite bond length changes, our algorithm still achieves high fidelity with the ground state.

\end{abstract}

\date{\today}
\maketitle
\end{CJK*}
\section{Introduction}
In recent years, quantum chemistry~\cite{szabo2012modern} has emerged as
a promising domain science field where quantum computers can potentially lead to a breakthrough~\cite{mcardle2020quantum}. Even for ground-state total energy problems, many systems exhibit strong character of many-body correlation effects (e.g., bond-stretching and transition states in chemical reactions), which are beyond the scope of widely used mean-field methods. As wave-function-based correlated methods remain computationally intractable against system size on classical computers, various quantum algorithms, 
such as iterative quantum phase estimation (iQPE)~\cite{aspuru2005simulated}, variational quantum eigensolver (VQE)~\cite{peruzzo2014variational,kandala2017hardware}, quantum adiabatic  evolution (QAE)~\cite{du2010nmr,Babbush2014}, and quantum annealing~\cite{xia2017electronic,streif2019solving}, have been proposed 
to tackle these quantum chemistry problems. Several quantum algorithms have been implemented and demonstrated for small molecules~\cite{mcardle2020quantum,lanyon2010towards,peruzzo2014variational,du2010nmr,kandala2017hardware,colless2018computation,streif2019solving,kandala2019error}. Further improved methods such as the equation of motion~\cite{ollitrault2019quantum} and the adaptive VQE~\cite{grimsley2019adaptive} have also been designed. Some of us have previously proposed an adiabatic method that uses a particular interpolation of two Hamiltonians~\cite{yu2021quantum}: (1) the final one being the targeted full many-body molecular Hamiltonian and (2) the initial one being the maximally commuting (MC) Hamiltonian constructed from the final Hamiltonian. Such a quantum adiabatic evolution from the initial MC Hamiltonian (MC-QAE)  approach works well for molecules near their equilibrium position, but the results become inaccurate at large atomic separations, due to energy crossing and the presence of dense low lying levels.  Previously we proposed a  quantum Zeno approach to overcome this issue, which uses projection to instantaneous eigenstates of the discretized time-dependent Hamiltonian~\cite{yu2021quantum}. However, the spectral projection in the quantum Zeno approach at the present is less practical  to implement than discretized Trotter evolution of the adiabatic Hamiltonian.

Here, we propose an adiabatic evolution method to compute ground state and low-lying excited state energies of the many-body Hamiltonian of molecular systems along an adiabatic geometric path, which we refer to as the geometric quantum adiabatic evolution (GeoQAE). The GeoQAE starts with an initial geometry where MC-QAE can solve the  ground state and low-lying excited states accurately. Then the system evolves  adiabatically following a smooth geometric deformation by, e.g., stretching or shrinking bonds and/or increasing or decreasing bond angles.  We show that the  GeoQAE approach does not suffer from the energy level acrossing issue in the MC-QAE approach at large atomic distances. 
We demonstrate its utility by computing the energies of the ground state and low-lying excited states of exemplary molecular systems, including H${}_2$O, CH${}_2$, and the potential energy surface of the chemical reaction of H${}_2$+D${}_2\rightarrow$ 2HD.

In this work, we found an important property of the MC Hamiltonian, proposed in our earlier work~\cite{yu2021quantum},  that in the qubit basis the MC Hamiltonian is equivalent to keeping only diagonal terms of the full configuration interaction (CI) Hamiltonian. The MC Hamiltonian in fact is closely related to the Hartree-Fock approximation, because under the fermionic basis it consists of diagonal one-body terms and two-body Hartree and Fock terms.  We show that the Hartree-Fock ground state is an eigenstate of the MC Hamiltonian and the Hartree-Fock ground-state energy is the corresponding eigenenergy.

The remaining of the paper is organized as follows. In Sec.~\ref{sec:molecular}, we briefly review the setup in the molecular Hamiltonian and discuss our choice of the initial Hamiltonian in the adiabatic evolution.  In Sec.~\ref{sec:GeoQAE}, we present our geometric adiabatic approach and describe the new protocol. There, we also present the analysis of the gap in the new adiabatic Hamiltonian to justify the approach. In Sec.~\ref{sec:results}, we show the results by the new GeoQAE approach applied to two molecules, H$_2$O and CH$_2$, yielding better results than the previous MC-QAE approach and the other quantum Zeno approach. In Sec.~\ref{sec:ChemReaction}, we present a study on a simplified chemical reaction of H$_2$+D$_2$ $\rightarrow$ 2HD, via the ground-state energy versus the bond lengths. In Sec.~\ref{sec:errors}, we discuss the errors and fidelity from the perspectives of the choice of the geometric path, the evolution time $T$ and the discrete step number $M$ in an evolution.  We make some concluding remarks in Sec.~\ref{sec:conclude}.
\section{Molecular Hamiltonian and the its maximally commuting Hamiltonian}\label{sec:molecular}

In this study, we focus on finding the ground-state energy and low lying excited states of a molecule at fixed atomic coordinates. For a given set of spin-orbitals,  the many-body Hamiltonian can be written in the second quantized form,
\begin{equation}
\label{eq:H}
	H=H^{(1)}+H^{(2)}=\sum_{i, j} t_{i j} a_{i}^{\dagger} a_{j}+\frac{1}{2} \sum_{i, j, k, l} u_{i j k l} a_{i}^{\dagger} a_{k}^{\dagger} a_{l} a_{j},
\end{equation}
where $i, j, k, l$  label the spin orbitals. The  one-body hopping $t_{ij}$ terms and two-body interacting terms $u_{ijkl}$ are given by the following expressions,
\begin{equation}
	\label{eq:spin-orbital}
	\begin{aligned}
	 t_{i j} &=\langle i|H^{(1)}| j\rangle\\
	 &\equiv\int d x_{1} \Psi_{i}\left(x_{1}\right)\left(-\frac{\nabla_{1}^{2}}{2}+\sum_{\alpha} \frac{Z_{I}}{\left|r_{1 I}\right|}\right) \Psi_{j}\left(x_{1}\right), \\ u_{i j k l} &=[ij\mid kl]\\
	 &\equiv\iint d x_{1} d x_{2} \Psi_{i}^{*}\left(x_{1}\right) \Psi_{j}\left(x_{1}\right) \frac{1}{\left|r_{12}\right|} \Psi_{k}^{*}\left(x_{2}\right) \Psi_{l}\left(x_{2}\right),
	\end{aligned}
\end{equation}
where $\Psi_{i}\left(x_{1}\right)$ are single particle spin wave functions of orbitals $i$. Note that $[ij|kl]$ is expressed in the so-called chemists' notation and the same quantity is denoted as $\langle ik|jl\rangle$ in the physicists' notation~\cite{szabo2012modern}.
These coefficients are calculated using the standard quantum chemistry package, {\tt PySCF}~\cite{PYSCF}, 
written in the {\tt Python} language. The choice of the basis set needs to be made to compute Eqs.~\ref{eq:spin-orbital}.  Large basis sets can in principle lead to better the numerical convergence with respect to the basis set size, however they require more qubits to implement the quantum algorithm. 
In this work the Slater-type orbital (STO)-3G basis (see e.g.~\cite{levine2009quantum,szabo2012modern}) is used as a compromise between accuracy and computational cost, while our main conclusion does not rely on the choice of the basis set.

The goal is to find the eigenstates and eigenenergies of Hamiltonian $H$ of molecules, i.e., $H|\psi_E\rangle= E|\psi_E\rangle$, in particular the ground state $\ket{\psi_G}$ and its energy $E_G$, as well as  low lying states.
In order for the problem to be solved on quantum computers,  we need to convert the above fermionic Hamiltonian to one composed of qubits. Existing methods include the Jordan-Wigner, parity, Bravyi-Kitaev, and superfast Bravyi-Kitaev transformations~\cite{seeley2012bravyi,setia2018bravyi,setia2019superfast}. By using any of these methods, we can transform the fermion operators into Pauli operators,
\begin{equation}
	\label{eq:Hp}
	H^{(P)}=\sum_i h_i P_i,
\end{equation}
where $P_i$'s are $n$-qubit Pauli operators and $h_i$'s are the corresponding coefficients. For the results presented below, we use the parity and  Jordan-Wigner transformation methods (see Appendix~\ref{appendix:mapping}).
\subsection{Initial Hamiltonian and initial state}
We will first use the Hatree-Fock procedure,  briefly reviewed in Appendix~\ref{app:HF}, which uses an initial set of atomic orbitals  with the full Hamiltonian, such as shown in Eq.~(\ref{eq:H}) and aims to obtain a new set of converged molecular orbitals, particularly indexed below by Greek letters $\alpha$, $\beta$, $\gamma$, etc. We will emphasize the molecular basis by adding the superscript MO to  the hopping and interaction coefficient in the resultant full Hamiltonian  as $t^{\text{MO}}$ and $u^{\text{MO}}$, respectively,
\begin{eqnarray}
\label{eq:Hfull}
	H_{\text{full}}^F&=&\sum_{\alpha,\beta} t_{\alpha \beta}^{\text{MO}} a_{\alpha}^{\dagger} a_{\beta}+\frac{1}{2} \sum_{\alpha, \beta,\gamma, \delta} u_{\alpha \beta \gamma \delta}^{\text{MO}} a_{\alpha}^{\dagger} a_{\gamma}^{\dagger} a_{\delta} a_{\beta},\nonumber
\end{eqnarray}
The resultant Hartree-Fock state is then a single-Slater determinant in this molecular basis.
For convenience, we denote the qubit version of $H_{\rm full}^F$ as $H_{\rm full}^P$, where the subscript $P$ means Pauli operators.

Our choice of the initial Hamiltonian $H_I^P$ (in the MC-QAE approach) is the diagonal part of the qubit version of the full Hamiltonian  $H_{\text{full}}^P$, which only consists of product of Z-terms and identities. For transformations mentioned above from the fermionic to the qubit representation, the corresponding eigenstates of $H_I^P$ are in the computational basis, which corresponds to single-determinant  states in the fermionic picture. In the fermionic Hamiltonian, one Slater determinant is the simultaneous eigenvector of the corresponding number operators $a_\alpha^\dagger a_\alpha$ for all $\alpha$. If the transformation converts $a_\alpha^\dagger a_\alpha$ into Pauli Z terms, and $a_\alpha^\dagger a_{\beta}$  non-Z terms for $\beta\ne\alpha$, then the diagonal part of $H^P_\text{full}$ is the summation of all possible combination of $a_\alpha^\dagger a_\alpha$, $a_\alpha^\dagger a_\beta^\dagger a_\beta  a_\alpha$ and $a_\alpha^\dagger a_\beta^\dagger a_\alpha  a_\beta$. 
Therefore, the fermionic version of our initial Hamiltonian is 
\begin{eqnarray}
	H_I^{F}&=&\sum_{\alpha} t_{\alpha \alpha}^{\text{MO}} a_{\alpha}^{\dagger} a_{\alpha}+\label{eq:HI}
	\\
	&&\frac{1}{2} \sum_{\alpha, \beta} (u_{\alpha \alpha \beta \beta}^{\text{MO}} a_{\alpha}^{\dagger} a_{\beta}^{\dagger} a_{\beta} a_{\alpha}-u_{\alpha \beta \beta \alpha}^{\text{MO}} a_{\alpha}^{\dagger} a_{\beta}^{\dagger} a_{\alpha} a_{\beta}),\nonumber
\end{eqnarray}
where the superscript $F$ is added explicitly  to emphasize the $H_I$ in terms of fermion operators. 
We remark that when written in the Hartree-Fock molecular basis,
the Hartree-Fock ground-state wave function is an eigenstate of $H_I^F$ with eigenenergy $E_{HF}$.
This can be easily verified by directly applying $H_I^{F}$ to the Hartree-Fock ground state $\ket{\psi_{HF}}=\ket{1..10..0}$ by filling up orbitals with lowest energies,
\begin{equation}
H_I^{F}\ket{\psi_{HF}}=E_{HF}\ket{\psi_{HF}},
\end{equation}
and obtaining  $E_{HF}$ to be the Hartree-Fock ground-state energy,
\begin{equation}
\label{eq:Ehf}
E_{H F}=\sum_{\alpha\in occ}\langle \alpha|H^{(1)}| \alpha\rangle+\frac{1}{2} \sum_{\alpha, \beta\in occ}([\alpha \alpha \mid \beta \beta]-[\alpha \beta \mid \beta \alpha]),
\end{equation}
where $[\alpha\beta \big\vert \gamma\delta]=u_{\alpha\beta\gamma\delta}^\text{MO}$ is the two-electron integral of spin-orbitals calculated in the Hartree-Fock molecular basis. More details are shown in the Appendix \ref{appendix:HF}.

We remark that the Hartree-Fock ground state is also an eigenstate of the so-called Fock operator  (a.k.a. the Hartree-Fock Hamiltonian) $F$  in Eq.~(\ref{eq:Fock}),but with an eigenenergy $\tilde{E}$ different from the Hartree-Fock ground-state energy~\cite{szabo2012modern}, due to  double counting in the interaction energy (i.e. a factor of two difference in the second term), 
\begin{equation}
\Tilde{E}=\sum_{\alpha\in occ}\langle \alpha|H^{(1)}| \alpha\rangle+ \sum_{\alpha, \beta\in occ}([\alpha \alpha \mid \beta \beta]-[\alpha \beta \mid \beta \alpha]).
\end{equation}

The Hamiltonian form~(\ref{eq:HI}), when converted to the qubit basis (denoted by $H_I$), is the one found   from the previous work~\cite{yu2021quantum}  constructed by finding the maximum commuting set of terms in the Hamiltonian. Prior results~\cite{yu2021quantum} show that for some molecules if one chooses a bond length near the equilibrium distance and uses the $H_I^P$ as the initial Hamiltonian, it can drive the Hartree-Fock initial state to the final ground state of the full Hamiltonian via adiabatic evolution  interpolating the two Hamiltonians,
\begin{equation}
	H(t)=\left(1-\frac{t}{T}\right)H_I^P+\frac{t}{T}H_{\text{full}}^P.
\end{equation}
\par
Below we shall see that the Hartree-Fock state does not always lead to the final ground state. However, we can choose other computational basis states or their superposition as the initial states. 
Note that the electron number is preserved during the evolution. So it is possible to choose an appropriate initial state to target the certain sectors with desired the electron number. 
In some particular cases where the system has additional symmetries and the direct evolution fails, we may use a symmetrized initial state instead to overcome the problem; see Sec.~\ref{sec:CH2} below. For example, supposing the MC Hamiltonian has two degenerate ground states $\ket{\psi_{HF}}\equiv\ket{\psi'_1}=\ket{01},\ket{\psi'_2}=\ket{10}$ and we have some prior knowledge that the final ground state is a singlet state, we can modify our initial state according to
\begin{equation}
\psi_{0}=\ket{\psi_{HF}} \quad \longrightarrow\quad \psi_{0}^{\prime}=\frac{1}{\sqrt{2}}(\ket{\psi'_1}-\ket{\psi'_2}).
\end{equation}
Due to the symmetry preserved by both initial and final Hamiltonians, choosing an initial state sharing the same symmetry as the expected finial ground state will restrict the evolution space within the subspace of desired  symmetry. Thus, the evolution will likely end up with the final ground state. We note that the above state $\psi'_0$ in the qubit basis can be initialized by a short-depth circuit.

\begin{figure*}[t]
	\centering
	\includegraphics[width=0.97\textwidth]{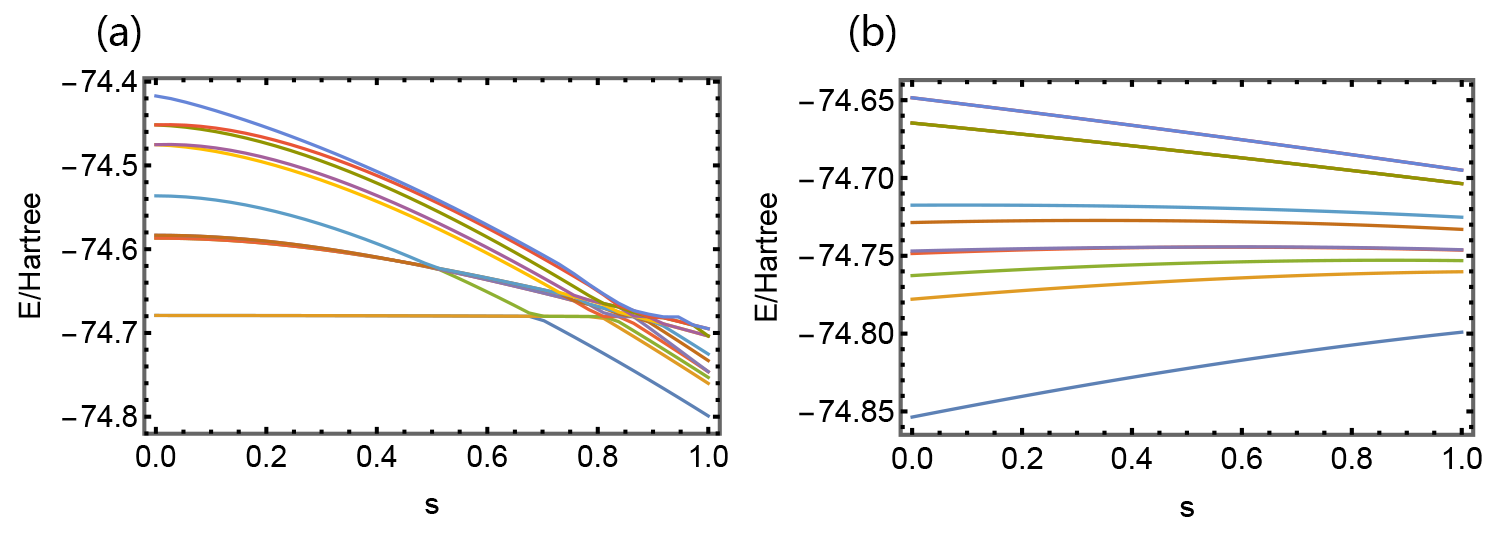}
	\caption{The lowest few energy levels of $H(s)$ for H$_2$O molecule at O-H distance $d=1.958\AA$. In case (a), the evolution begins with maximum commuting Hamiltonian at $d=1.958\AA$, see Eq.~\ref{eq:H1}; in case (b), the evolution begins the full Hamiltonian at $d=1.758\AA$ and ends at the one at $d=1.958\AA$; see Eq.~\ref{eq:H2}.}
	\label{H2Ogap}
\end{figure*}
\section{Geometric adiabatic path and the new protocol}\label{sec:GeoQAE}

To overcome that problem that the  MC-QAE approach  fails at large bond lengths despite that it works well near the equilibrium distance,  we propose a new idea, justify it and examine it in several examples. Suppose we have the qubit Hamiltonian $H_P(\bm{r}_0)$ (and have obtained its ground state, e.g. using the previous MC-QAE method that works) for a molecule at $\bm{r}_0$ near the equilibrium position. In order to obtain  the ground state and energy of $H^P_\text{full}(\bm{r})$ at large molecular distances, we construct a series of subsequent Hamiltonians $H^P_\text{full}(\bm{r}_1), H^P_\text{full}(\bm{r}_2),..., H^P_\text{full}(\bm{r}_N=\bm{r})$ so that the sequence $\bm{r}_0$, $\bm{r}_1$, \dots, $\bm{r}_N$ represents a gradual change in the atomic positions. Then we  let the system evolve piece-wise according to their interpolation as follows,
\begin{eqnarray}
\label{GeoEq}
	\nonumber
	H_0(t)&=&\left(1-\frac{t}{T}\right)H_I^P(\bm{r}_0)+\frac{t}{T}H_{\text{full}}^P(\bm{r}_0),\\\nonumber
	H_1(t)&=&\left(1-\frac{t}{T}\right)H_{\text{full}}^P(\bm{r}_0)+\frac{t}{T}H_{\text{full}}^P(\bm{r}_1),\\\nonumber
	\vdots\\
	H_N(t)&=&\left(1-\frac{t}{T}\right)H^P_{\rm full}(\bm{r}_{N-1})+\frac{t}{T}H^P_{\rm full}(\bm{r}_N).
\end{eqnarray}
For each step, the evolution from $H_P(\bm{r}_i)$ to $H_P(\bm{r}_{i+1})$ is likely to succeed due to the continuity with $\bm{r}$ of the Hamiltonian $H_P(\bm{r})$. 

\subsection{Gap analysis}
To justify and illustrate that this alternative approach  works, we compare the spectrum of two path dependent Hamiltonians leading to the same final Hamiltonian $H_P(r=1.958\AA)$ for the water molecule H${}_2$O. 
The first one begins with the MC Hamiltonian $H_I(r=1.958\AA)$ associated with the final Hamiltonian $H_P(r=1.958\AA)$,
\begin{equation}H_1(s)=(1-s) H_{I}^P( r=1.958\AA)+ s H_{\text{full}}^{P}( r=1.958\AA).
\label{eq:H1}
\end{equation}The second Hamtilonian
begins with the full Hamiltonian at an earlier distance, $H_P(r=1.758\AA)$ and ends with the final Hamiltonian $H_P(r=1.958\AA)$,
\begin{equation}H_2(s)=(1-s) H_{\text{full}}^{P}( r=1.758\AA)+ s H_{\text{full}}^{P}( r=1.958\AA).
\label{eq:H2}
\end{equation}
 Their spectra as a function of $s$ are shown in Fig.~\ref{H2Ogap}. There are energy crossings in $H_1(s)$, as shown in Fig.~\ref{H2Ogap}a, and this is the reason that the direct MC-QAE fails. However, assuming one can arrive at the ground state of a Hamiltonian at, e.g., a shorter distance $r=1.758\AA$,  one can use this Hamiltonian as the initial one to adiabatically evolve its ground state (as well as excited states) to that of $H_{\text{full}}^{P}(r=1.958\AA)$ or one at a shorter distance. The  resultant path-dependent Hamiltonian $H_2(s)$ has   energy levels  smoothly connected without any crossing, as shown in Fig.~\ref{H2Ogap}b. This is perhaps expected, as it corresponds to a small stretching of the chemical bonds. Thus, we see that     the energy-crossing problem is solved by evolving from the nearby bond length $r=1.758\AA$ instead of the maximum commuting Hamiltonian at $r=1.958\AA$.

The above is not an isolated example. In fact,
for most cases, the energy gaps between levels during the evolution from $H_{\text{full}}^P(r_i)$ to $H_{\text{full}}^P(r_{i+1})$ remain finite and large, even in the region of large bond lengths where direct evolution (from the MC Hamiltonian) fails. This consequence relies on the continuity in the Hamiltonians when atoms are displaced gradually in changing their geometry. (See  Appendix~\ref{continuity} for several methods we use to ensure continuity in the Hamiltonians.) For the cases where the Hartree-Fock method converges, the changing of one- and two-body coefficients is small when $\Delta \bm{r}=\bm{r}_{i+1}-\bm{r}_{i}$ is sufficiently small.  This results in the separation of energy levels in the interpolated Hamiltonian. In our calculations, we found $|\Delta \bm{r}|\approx 0.1 r_0$ is good enough for the evolving Hamiltonian to maintain  gaps. However, when classical Hartree-Fock fails to converge, we may need to transform the Hamiltonian $H_{\text{full}}^P(\bm{r}_{i+1})$ back to the previous molecular basis $\{\ket{\varphi_\mu}_{\bm{r}_i}\}$, in order to maintain numerical continuity of the coefficients. (As also seen below in Sec.~\ref{sec:errors}, despite the small gap when $\Delta r$ is not small, the fidelity of the final ground state can still be large.) In our experience, by using the converged solution in a prior distance as the initial guess in the (classical) Hartree-Fock procedure usually results in the convergence, even if the random initial guess does not.
\par
\subsection{New protocol}

In the following we assume that the Hamiltonians are all written in the converged Hartree-Fock molecular basis. With the assumption that the changing of the geometric configuration $\Delta \bm{r}_i=\bm{r}_{i+1}-\bm{r}_i$ is much smaller than the atomic distance, we can ensure that the next Hamiltonian is not dramatically different from the previous one. We refer to this approach as the geometric quantum adiabatic evolution (GeoQAE) and the whole protocol can be summarized as follows.
\begin{quote}
	1. Choose a proper bond length $r_0$ (near the equilibrium distance),  construct a Hamiltonian of a molecule, and transform the  fermion operators to Pauli operators;\\
	2. Choose the all-Z terms to construct the initial Hamiltonian $H_I^P(r_0)$ and the Hartree-Fock ground state or other appropriate computational states as the initial state $\psi_{0}$;\\
	3. Discretize the time steps 
	by $t_k=\frac{k}{M}T$ and obtain a series of Hamiltonians $H_0(t_k)$ in Eq.~\ref{GeoEq};\\
	4. Perform the  evolution operator $e^{-i H_0(t_k)\Delta T}$ on $\psi_{0}$ successively for $k=1,\dots,M$, to aim at one eigenstate of the final Hamiltonian $H_{\text{full}}^P(r_0)$;\\
	5. Choose a suitable series of bond lengths $r_1,r_2,...,r_N$ with $r_N$ equal to the desired bond length. Then repeat similar evolution procedure (3 and 4) with the initial Hamiltonian $H_{\text{full}}^P(r_i)$ and final Hamiltonian $H_{\text{full}}^P(r_{i+1})$ multiple times.  We expect to obtain an approximated ground eigenstate of $H_{\text{full}}^P(r_N)$. We can perform measurements to obtain its corresponding energy.
\end{quote}
We remark that the above points 1-4 are exactly the MC-QAE approach, and the protocol for GeoQAE uses  MC-QAE as its first step at an appropriate molecular distance $r_0$ (usually around the equilibrium position, e.g. such as determined the Hartree-Fock method)  to arrive at the exact molecular eigenstates  and subsequently undergoes the  geometric change of the molecular configuration, as in point 5 above to reach the corresponding eigenstates. 
\section{Results on molecular energies and fidelities}
\label{sec:results}
We apply our geometric adiabatic path approach to three different systems $\mathrm{H}_2\mathrm{O}$, CH$_2$, and a simple chemical reaction $H_2+D_2\longrightarrow 2HD$. We numerically calculate the following evolution, 
\begin{equation}
\label{eq:Adiabatic}
\ket{\psi(T)}\approx e^{-i H(T)\Delta T}e^{-i H(T-\Delta T)\Delta T}\dots e^{-i H(\Delta T)\Delta T}\ket{\psi(0)},
\end{equation}
and evaluate the resultant energy, where $\Delta T=T/M$ and $|\psi(0)\rangle$ is a suitably chosen initial state.  The molecular Hamiltonian (i.e. its coefficients $t$'s and $u$'s) is calculated by the {\tt PySCF} package~\cite{PYSCF} and transformed to Pauli operators by the {\tt Qiskit}~\cite{Qiskit}. The transformation we employ in this work includes the Jordan-Wigner transformation and the parity mapping. We also freeze the $1s$ orbital of larger atoms, such as C and O. If not specified, we use the Hartree-Fock state as our initial state. The case of CH${}_2$ requires the use of other initial states. We also compare our results by the GeoQAE method with those obtained  with direct evolution (i.e. the MC-QAE) from $H_I^P(r_i)$ to $H_{\text{full}}^P(r_i)$.

We now present several case studies.
\subsection{H${}_2$O}
\begin{figure}[!h]
	\centering
	\includegraphics[width=0.42\textwidth]{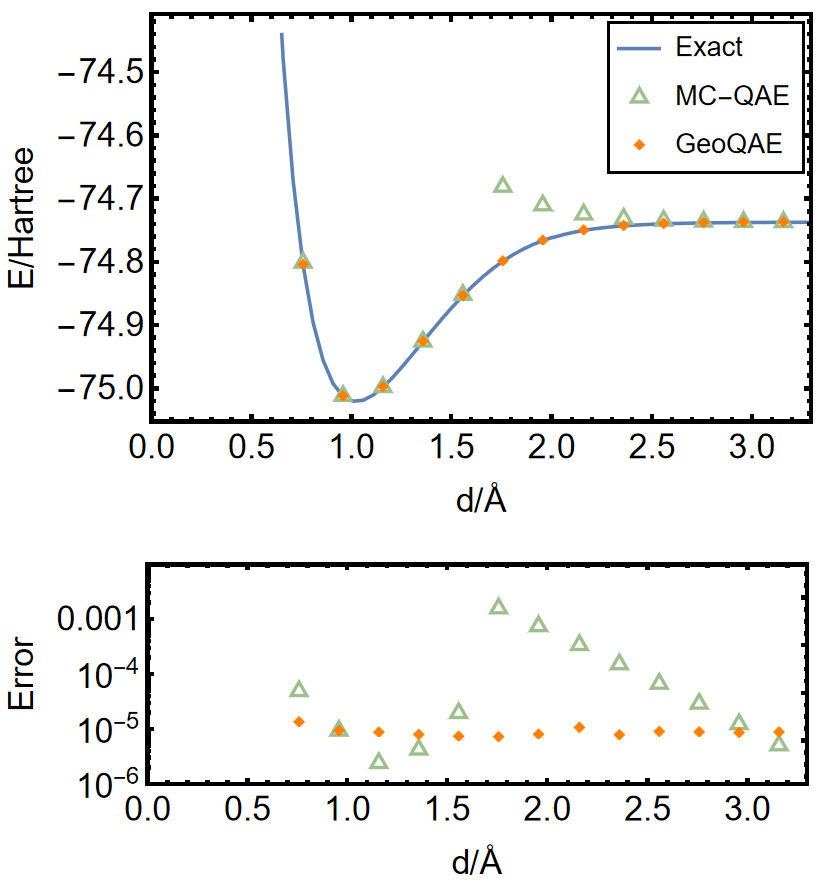}
	\caption{Top panel: The ground-state energy of H$_2$O vs. the distance $d$ between the O and H atoms with  different approaches: classically exact diagonalization (the solid curve), the MC-QAE (the triangles) and the GeoQAE (the dots). The H-O-H angle is fixed to the equilibrium angle $\theta=104.45^{\circ}$. The bottom panel shows the errors in targeting the ground-state energies using the MC-QAE and GeoQAE. We use $T=40$ and $M=20$ for both methods. The latter (GeoQAE) may have larger errors when $T$ and $M$ are small, but the error  can be improved by increasing the two parameters.}
	\label{H2O}
\end{figure}

\smallskip\noindent {\bf Ground-state energy}. For $\text{H}_2$O molecule, we fix the H-O-H angle to be the equilibrium angle $\theta=104.45^{\circ}$ and will seek the ground-state energy of the molecule as we change the O-H bond. The parity mapping method along with the qubit reduction and freezing of 1s orbital of O atom are used to reduce and simplify the Hamiltonian. The final qubit number to represent the Hamiltonian is 10.

\begin{figure}[!h]
	\centering
	\includegraphics[width=0.42\textwidth]{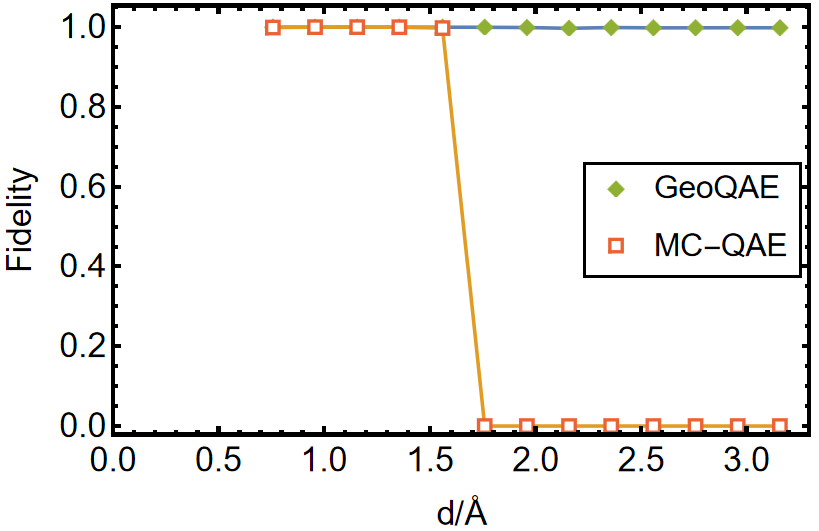}
	\caption{The results of the fidelity of H${}_2$O. The H-O-H angle is fixed to the equilibrium angle $\theta=104.45^{\circ}$.  We set $T=40$ and $M=20$ for both methods.}
	\label{h2oF}
\end{figure}

In our GeoQAE calculation, we choose $r_0=0.9584\AA$ in the initial MC-QAE step with the Hartree-Fock state as the initial state to obtain the ground state of the molecular Hamiltonian at this position. Then for other positions, we follow the procedure outlined in Sec.~\ref{sec:GeoQAE}.  
Compared to the direct evolution from the MC Hamiltonian, the GeoQAE method gives  very good results (e.g. with relative errors being $10^{-5}$ or smaller) even in the large atomic distances where MC-QAE fails, as shown in Fig.~\ref{H2O}. Nevertheless, due to the sequential evolution the error will accumulate from all steps of evolution. Therefore, to achieve the same fixed accuracy at larger atomic distances, we may need more discretization steps $M$ and a larger evolution time $T$. But a recent study shows that the accumulation error in using Trotterization for the adiabatic evolution is not as severe as one would expect if the initial state is an eigenstate~\cite{yi2021spectral}, as we have also observed in the bottom panel of Fig.~\ref{H2O}. 
\begin{figure}[t]
	\centering
	\includegraphics[width=0.42\textwidth]{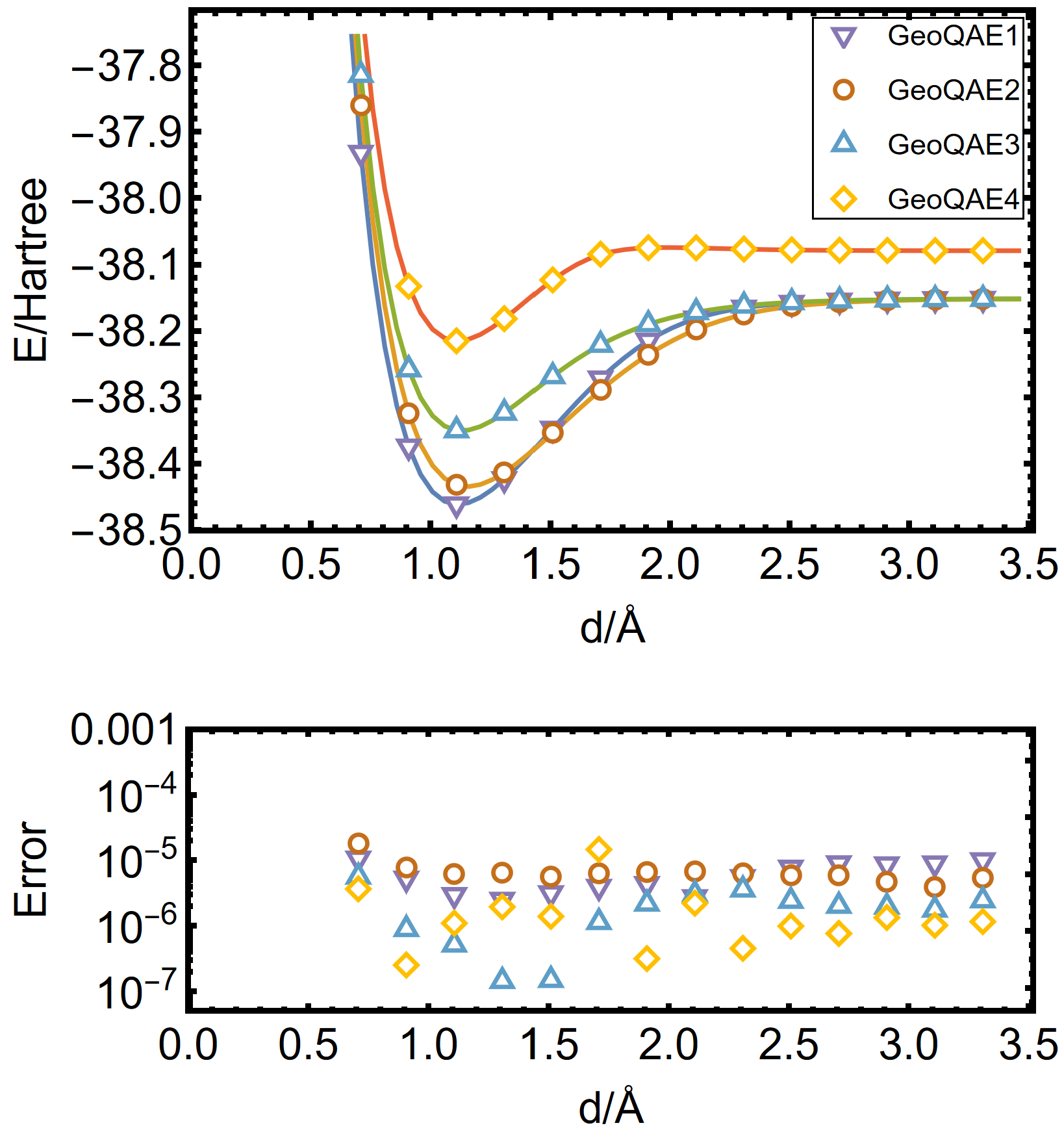}
	\caption{The ground state energy of CH$_2$ vs. the distance $d$ between the C and H atoms with different initial states. The H-C-H angle is fixed to the equilibrium angle $\theta=101.89^{\circ}$. We choose $T=60$ and $M=30$ for all the cases. Top panel: The curves represent exact solutions from directly diagonalizing the Hamiltonian and the four different legends (labeled by GeoQAE1-4) represent results from simulated evolution on four different initial states (see the main text). The bottom panel displays the respective errors in targeting the eigenenergies.}
	\label{CH2}
\end{figure}

\smallskip
\noindent {\bf Ground-state fidelity}.
We also  compute the fidelity of the final evolved state $\ket{\psi(T)}$ with  the exact molecular ground state $\ket{\psi_g}$ of $H_P$, i.e. $f(\psi_g,\psi(T))=|\langle \psi_g|\psi(T)\rangle|$~\cite{nielsen2002quantum}. In these simulations we use evolution time $T=40$  and discretize the continuous time evolution to $M=20$ time slices. 
We compare the resultant  ground-state fidelity using two different evolutions, one with MC-QAE and the other with GeoQAE. 

Even though the energy calculation by the MC-QAE (see Fig.~\ref{H2O}) may only has at most 0.1\% error in the range we consider, it is actually not targeting the ground state (for $r\gtrsim 1.5\AA$), but an excited state (or some combination of them) instead. As the result of the fidelity in Fig.~\ref{h2oF} shows, the final state from MC-QAE at large $r$'s has zero fidelity with the ground state. On the other hand the evolved final state via GeoQAE does have close to unity fidelity with the ground state of the final Hamiltonian.
 
With the MC-QAE, for $d\gtrsim1.76\AA$, the fidelity goes to nearly 0. This is because the direct adiabatic evolution from the MC Hamiltonian experiences level crossings and can only find excited states for large molecular distances $d$, which was noticed in our previous work~\cite{yu2021quantum} and illustrated in Fig.~\ref{H2Ogap}a. But with the geometric evolution, we can maintain the fidelity to nearly 1 even at large atomic distances. Alternatively, one may also explore other nearby  atomic arrangement (to the target distance) as the starting point of the GeoQAE algorithm.

\subsection{CH$ _2 $}
\label{sec:CH2}

\smallskip \noindent {\bf Energy of the lowest few levels and the choice of initial states}.
For the $\text{C}\text{H}_2$ molecule, we fix the H-C-H angle to be the equilibrium angle $\theta=101.89^{\circ}$ and vary the C-H bond. The results of simulating our GeoQAE method are shown in Fig.~\ref{CH2}.  Here we freeze the 1s orbital of the C atom, similar to the $\text{H}_2$O case,  but use the Jordan-Wigner mapping to  qubits.
The final qubit number to represent the Hamiltonian is 12, as we do not have the qubit reduction as in the parity mapping.

\begin{figure}[t]
	\centering
	\includegraphics[width=0.42\textwidth]{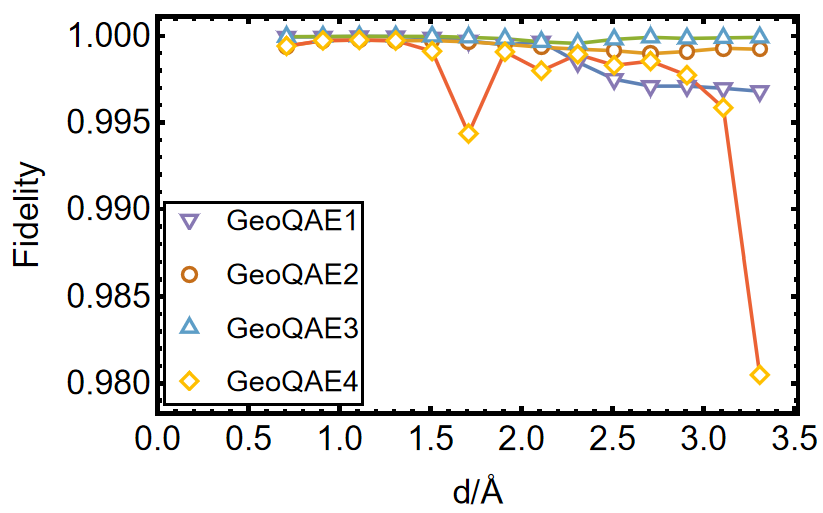}
	\caption{The results of the fidelity of CH${_2}$ vs. the distance $d$ between the C and H atoms with different initial states. The H-C-H angle is fixed to the equilibrium angle $\theta=101.89^{\circ}$. We choose $T=60$ and $M=30$ for all the cases.}
	\label{ch2F}
\end{figure}

\begin{figure*}[t]
	\centering
	\includegraphics[width=0.86\textwidth]{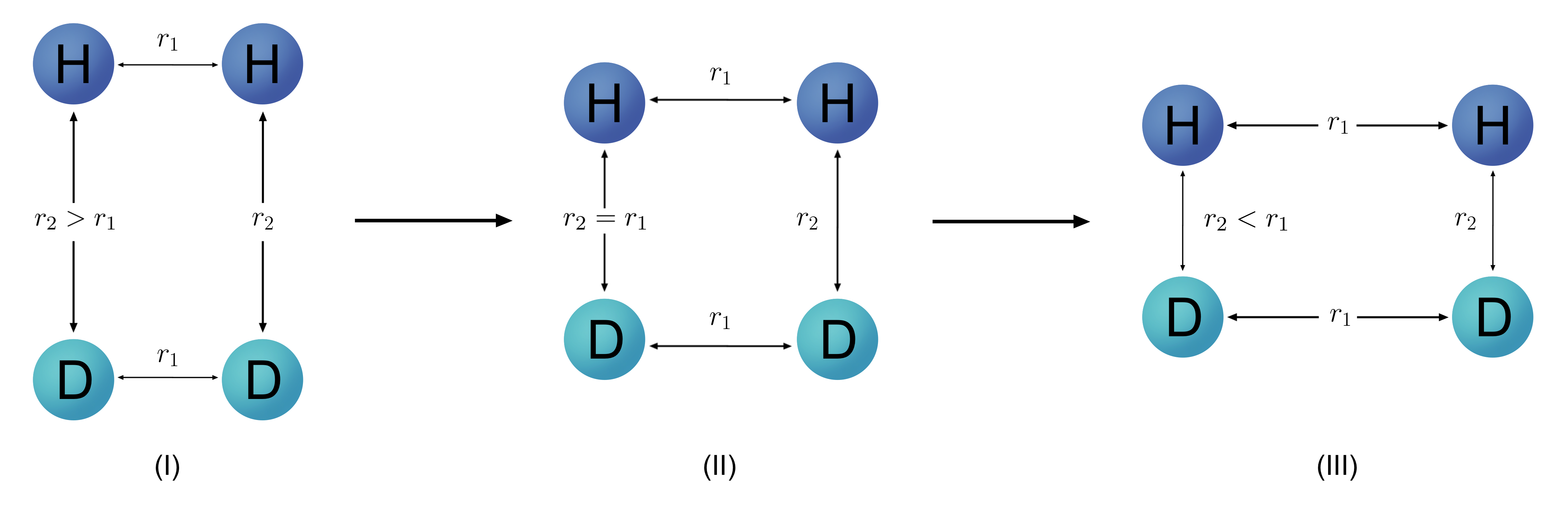}
	\caption{The geometric configurations for chemical reaction $H_2+D_2\longrightarrow 2HD$.}
	\label{h2d2}
\end{figure*}
We list the lowest 6  eigenstates of the MC Hamiltonian $H_I^P(\bm{r}_0)$ at $r_0=1.1089 \AA$,
\begin{eqnarray*}
 &&   \phi_1=\ket{110000,111100}, \ E_1=-10.5403\\
 &&   \phi_2=\ket{111100,110000}, \ E_2=E_1\\
 &&   \phi_3=\ket{110100,110100}, \ E_3=-10.4989\\
&&    \phi_4=\ket{110100,111000}, \ E_4=-10.4868\\
 &&   \phi_5=\ket{111000,110100}, \ E_5=E_4\\
 &&   \phi_6=\ket{111000,111000}, \ E_6=-10.3105,
\end{eqnarray*}
where the energy values (in the unit of Hartree) are calculated after orbital-freezing. We freeze the 1s orbital of the carbon atom, so the number of the orbitals is reduced to 12 (including spins) and the number of electrons is reduced to 6. This gives an extra energy shift along with the nuclear repulsion energy $\Delta E=-27.8618$ (Hartree).
In the above, the first 6 binary numbers in the kets represent the occupation of the molecular orbitals with spin up (usually labeled as $\alpha$) in the order of increasing energy and the last 6 numbers represent the occupation of the same set of molecular orbitals but with spin down (usually labeled as $\beta$). 
The Hartree-Fock wave function turns out to be $\phi_6$, which is not the lowest eigenstate of the $H_I^P$. The wave functions $\phi_1$ and $\phi_2$ are triplets and degenerate ground states (w.r.t. $H_I^P$). The wave functions $\phi_4$ and $\phi_5$ are also degenerate w.r.t. $H_I^P$ but they differ by opposite spins in the last two occupied orbitals, i.e. $\downarrow_3\uparrow_4$ vs. $\uparrow_3\downarrow_4$ (where the subscripts indicate the molecular orbitals). 
Due to the fact that the MC and full Hamiltonians  conserve the total spin angular momentum and its z component, we can consider combinations of two single-Slater determinants (having the same initial energy)  to form singlet and/or triplet states of opposite spins as the initial states. 
For example,  from $\phi_4$ and $\phi_5$  we can define a singlet $\psi_{3}\equiv(\phi_4-\phi_5)/\sqrt{2}$ and a triplet $\psi_{1,c}\equiv(\phi_4+\phi_5)/\sqrt{2}$ wave functions for the initial states. For convenience, we also define $\psi_{1,a}\equiv\phi_1$, $\psi_{1,b}\equiv\phi_2$,
$\psi_2\equiv\phi_3$, and $\psi_4\equiv\phi_6$.

The reason to define these four sets of wave functions $\psi_1$'s, $\psi_2$, $\psi_3$ and $\psi_4$ is that their evolution under the combination of $(1-s)H_I^P + s H_{\text{full}}^P$ gives rise to the lowest four energy levels of the final Hamiltonian around the equilibrium position  (with the ground-state degeneracy being 3). We have  carried out this step of evolution (which is the MC-QAE procedure) numerically to verify that they indeed achieve the corresponding energies and levels with $99\%$ fidelity or above.  Subsequently, we proceed with the GeoQAE steps to obtain energies and wave functions at other molecular distances (both smaller and larger). Shown in Fig.~\ref{CH2}, the three energy curves originating from $\psi_1$'s, $\psi_2$ and $\psi_3$ remain the lowest three energy curves, whereas the curve originating from $\psi_4$ evolves to a higher excited state at large molecular distances (as seen from its gap between the three lower curves). It is interesting that the lowest two energy curves cross, i.e. the three-fold degenerate ground states become the first excited states at larger molecular distances. Given that these different levels have different symmetries, the adiabatic evolution will not lead to any mixing.

We remark that the superposition states, such as $\psi_{1,c}$ and $\psi_3$, are similar to the Greenberger-Horne-Zeilinger states and can be easily created and initialized by a short-depth circuit consisting of Hadamard and CNOT gates.
In the above, $\phi_i$'s  were calculated by direct diagonalizing the initial MC Hamiltonian, which is essentially classical (with Hamiltonian terms being products of Pauli Z and identity operators). Despite that finding the lowest configuration of a generic classical (e.g. Ising spin-glass) Hamiltonian can be NP,  the time complexity of finding its low-lying states is still much smaller than those of the final quantum Hamiltonian for  molecules. However, it is also not known whether the MC Hamiltonians  from quantum chemistry problems are necessarily NP-hard.

\smallskip
\noindent {\bf Fidelity with lowest three levels and the other excited level}. In addition to the energy curves, we also calculate the fidelity of the evolved states $\Psi_T$ with the  corresponding eigenstates $\Psi_E$ (exactly solved numerically), $f(\psi_T,\psi_E)=|\langle\psi_E|\psi_T\rangle|$, and the  results are shown in Fig.~\ref{ch2F}. Despite that the ground states (near equilibrium) are three-fold degenerate, their spin configurations are different and hence their states can be numerically separated and distinguished, the use of the fidelity expression $f(\psi_T,\psi_E)=|\langle\psi_E|\psi_T\rangle|$ is appropriate. In case that the numerically solved degenerate states $\Psi_{E,i}$ can be arbitrary superposed, one can average over the degeneracy by using $f\equiv\sqrt{\sum_{i=1}^{d_g}|\langle \Psi_{E,i}|\psi_{T}\rangle|^2}$, assuming $d_g$ degenerate states.

We can conclude from the fidelity results shown in Fig.~\ref{ch2F} that the geometric evolution can find all the lowest three levels and the other excited level (the latter corresponding to the 4th level around the equilibrium) with high fidelity. 
In contrast, the direct evolution from Maximum commuting Hamiltonian (i.e. MC-QAE) encounters problem at larger distances (not shown explicitly)~\cite{yu2021quantum}, similar to the case of H$_2$O in the previous section.

\section{Chemical reaction}\label{sec:ChemReaction}

The knowledge of the potential energy surface of a chemical reaction is critical to understand the reaction dynamics and kinetics. In particular, reaction energies and reaction barrier heights (i.e., the energy differences between reactants/products and the transition state) are key quantities that dictate the reaction energetics. It is important to develop quantum algorithms to compute the reaction potential energy surface.

In the previous section, we have seen that molecular energies can be accurately obtained by our GeoQAE, which extends the validity of the MC-QAE approach to a much greater extent. As a further application, we now apply this method to obtain the energy landscape of one chemical reaction. We use $H_2+D_2\longrightarrow 2HD$ as an example. Such a system is also of interest because of experimental realization of cold controlled chemistry~\cite{balakrishnan2016perspective}.

\begin{figure}[!h]
	\centering
	(a) \includegraphics[width=0.48\textwidth]{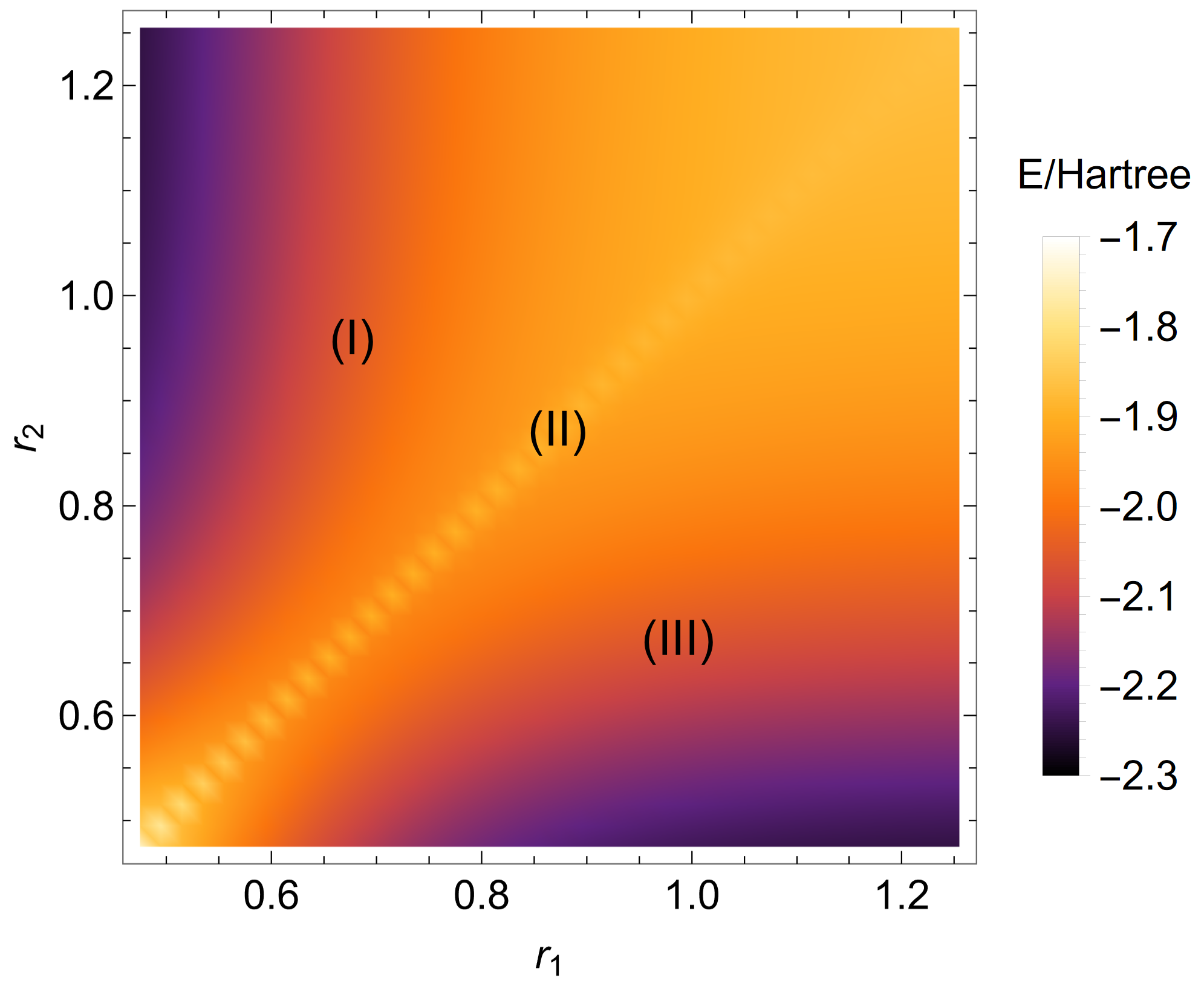}
    (b)	\includegraphics[width=0.48\textwidth]{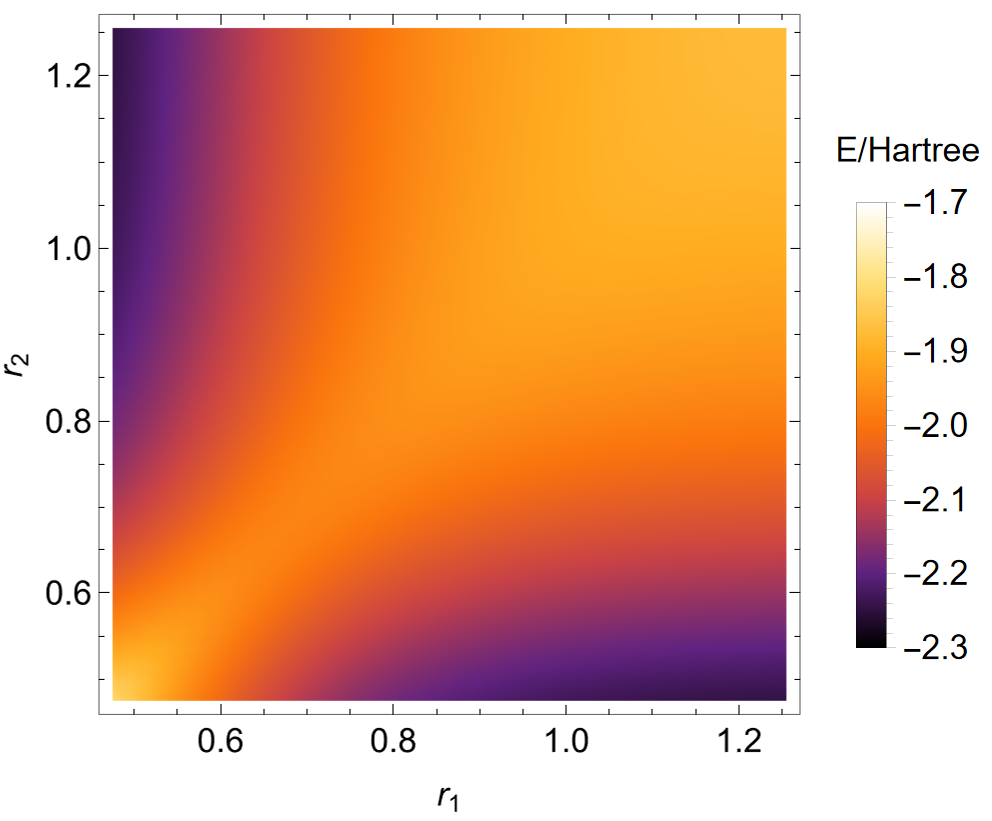}
	\caption{(a) The results of the potential energy surface using MC-QAE. (b) The results of the potential energy surface after the correction from geometric adiabatic path. We choose $T=80$ and $M=40$ for both methods. The correction only takes place in the region between the two dashed line in fig.~\ref{h4err}.}
	\label{h42dB}

\end{figure}

To reduce  the complexity of the problem, we  consider that the four atoms are initially set as the vertices of a rectangle with smaller horizontal edge; see Fig.~\ref{h2d2}. And the reaction is simplified to a process that the horizontal edge is getting longer, and it  can be  viewed effectively as the horizontal hydrogen molecules break up and form two new vertical molecules separately each with one deuterium. The highest energy of such process is when the four atoms become a square.  Thus, in order to know the energy barrier of such reaction, it is crucial to know the accurate energy of the square configuration~\cite{gasperich2017h4}.  For those cases that are not square, we find that the MC-QAE method works fairly well. But when it comes close to a square configuration, the MC-QAE fails because of the energy crossing arising from the extra symmetry of a square. In this symmetric regime, the GeoQAE performs much better, as demonstrated in  Figs.~\ref{h42dB}a and~\ref{h42dB}b.\par

\begin{figure}[!h]
	\centering
	\includegraphics[width=0.48\textwidth]{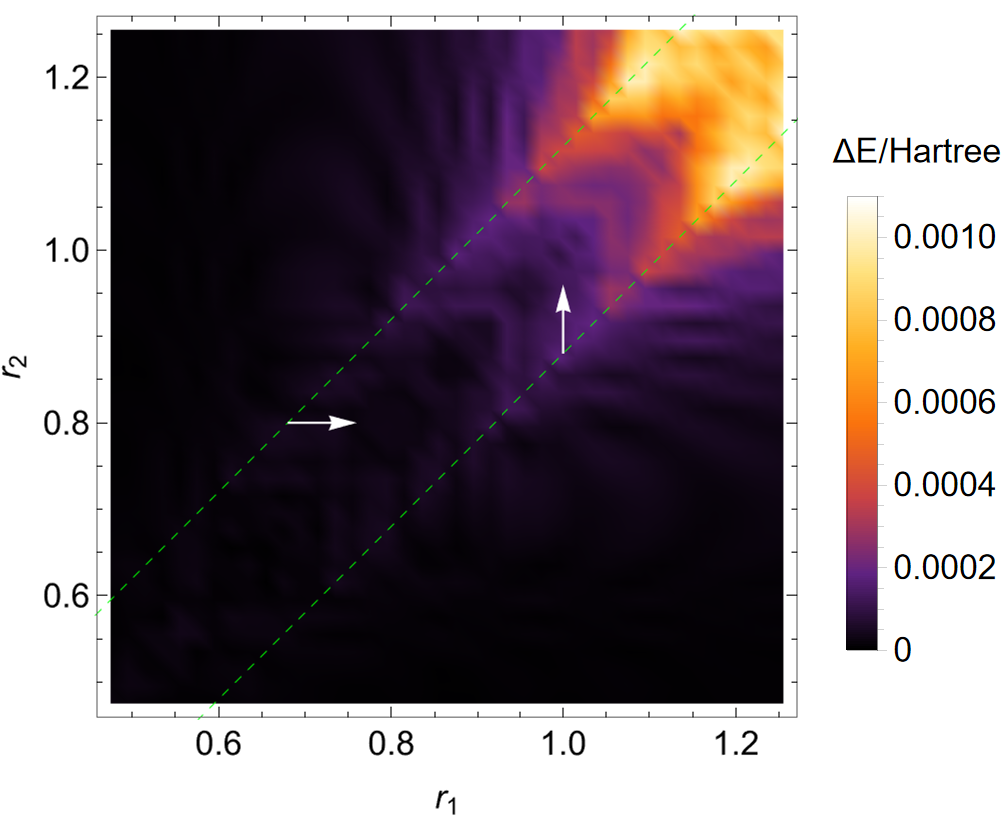}
	\caption{The energy difference compared to the exact results of the potential energy. The points within the two dashed line are calculated by GeoQAE evolving from the nearest (in the sense of keeping either $r_1$ or $r_2$ the same) boundary of this region. And the white two arrow shows the two possible evolution directions. The points outside the dashed line is calculated directly from MC-QAE.}
	\label{h4err}
\end{figure}
\begin{figure}[!h]
	\centering
	\includegraphics[width=0.42\textwidth]{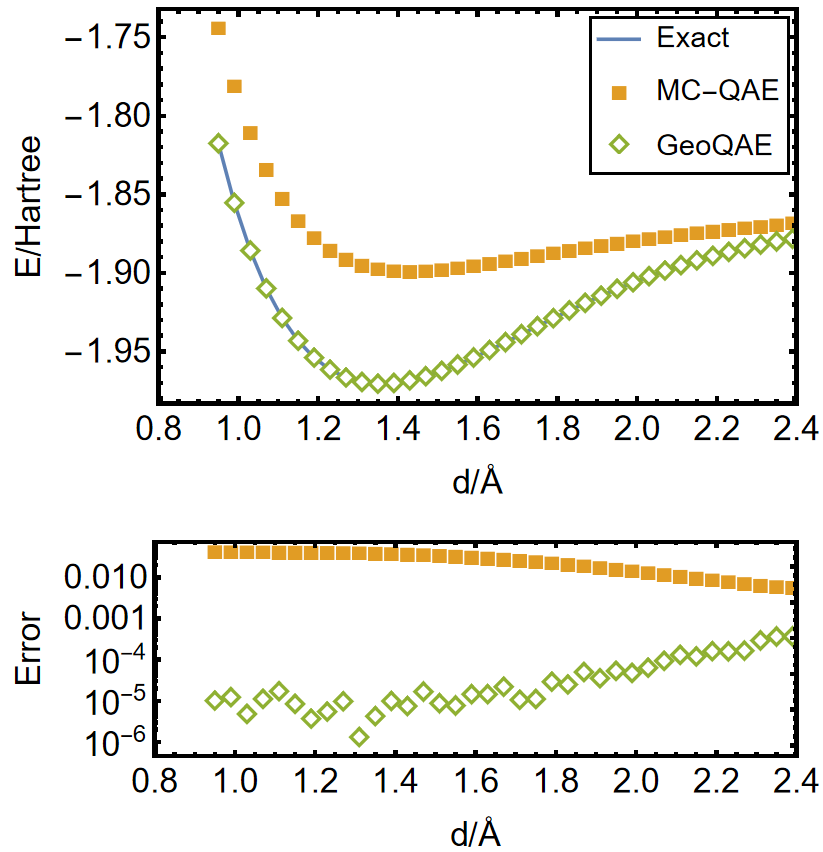}
	\caption{Upper panel: The results of the potential energy along the $r_1=r_2$ line for two different methods, MC-QAE and GeoQAE. The lower panel shows the relative errors from the two methods.}
	\label{h4}
\end{figure}
In our calculations, we use the STO-3G basis for orbitals and the Jordan-Wigner mapping to convert the fermionic Hamiltonians to the corresponding qubit Hamiltonians. The total qubit number required is 8. For the off-diagonal points representing unequal distances in the two graphs, we use the MC-QAE for both cases. For the diagonal points, we use a two-step  approach: first we let the system evolve to its nearest off-diagonal neighbor (via the MC-QAE), then do an adiabatic evolution to evolve to its ground state at the diagonal point via the GeoQAE, which is illustrated by the arrows in Fig.~\ref{h4err}. For each run of evolution from one distance to another, we set $T=80$ and $M=40$.

The energy difference between the GeoQAE and the exact solution is shown Fig.~\ref{h4err} and the results are very accurate except in the region of larger $r$'s, which can be improved by using larger $M$ and $T$ (see e.g. Sec~\ref{sec:errors}). In Fig.~\ref{h4}, we explicitly compare the results of the MC-QAE and GeoQAE at $r_1=r_2$, i.e. the square configuration. It can be seen that the energies from the MC-QAE are visibly higher than those from the GeoQAE, with the latter having $10^{-4}$ relative errors or smaller. This again shows the drastic improvement of the new approach by the GeoQAE where the MC-QAE fails.

\section{Error analysis and computational complexity}
\label{sec:errors}
Apart from the possible noise and error from real quantum computers, which was discussed in our previous paper~\cite{yu2021quantum}, the theoretical error of the geometric adiabatic path mainly depends on three parts: the choice of geometric path, the evolution time $T$ and the discretizing step number $M$. We discuss these below. 
\subsection{Choices of geometric path}
The geometric path is crucial for the whole procedure. While in principle when $\Delta \bm{r}$ is small enough, the physical evolution is continuous, it is still important in practice to choose a proper $\Delta r$ that is neither too large that the adiabatic evolution fails nor too small that the computational time becomes too long and errors accumulate too much. 
\par
\begin{figure}[!ht]
	\centering
	\includegraphics[width=0.46\textwidth]{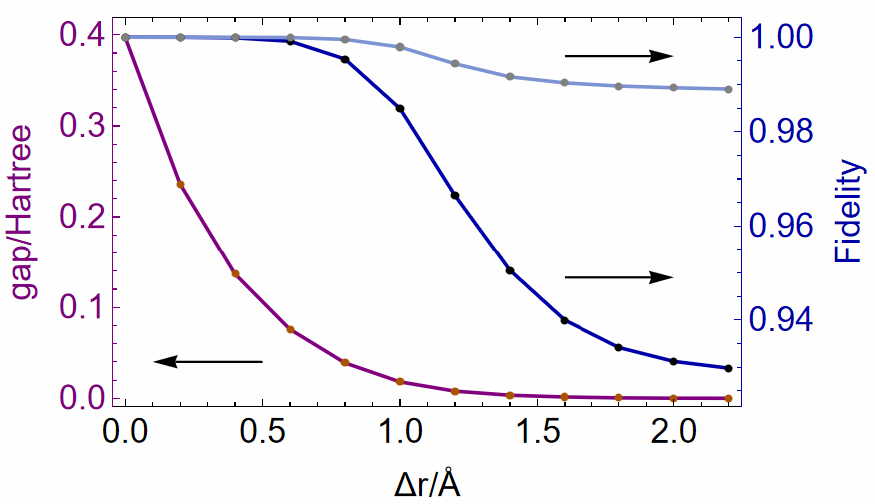}
	\caption{The results for the energy gaps of \textit{one-step} GeoQAE evolution and the fidelity of the ground-state wave functions for H$_2$O vs. different $\Delta r$. The evolution is from $H_{\text{full}}^P(r_0)$ to $H_{\text{full}}^P(r_0+\Delta r)$, with initial O-H bond length $r_0=0.9584 \AA$ and H-O-H angle fixed $\theta=104.45^{\circ}$. The gap minimum between the ground state and the first excited state for $H(s)=(1-s)H^P_\text{full}(r_0)+s H^P_\text{full}(r_0+\Delta r)$ shows a trend of exponential decay with $\Delta r$. The upper fidelity curve is obtained with $T=160$ and $M=80$ and the lower fidelity curve is obtained with $T=40$ and $M=20$. There is a substantial improvement using larger $T$ and $M$ values.}
	\label{h2oDf}
\end{figure}

Here we explore the length of $\Delta r$ as an example. When $\Delta r$ gets larger, it requires fewer (accumulated) steps to evolve to our desired bond length. However, within each evolution from one distance to another, the minimum gap will get smaller. Thus, we need more evolution time for each step in order to achieve the same accuracy. As shown in Fig.~\ref{h2oDf}, the energy gaps decrease exponentially with $\Delta r$ as expected. Despite this,  the ground-state fidelity can still achieve a value of 0.93 with $\Delta r=2.0\AA$ with $T=40$ and $M=20$. This can be further improved to 0.99 larger parameter values $T=160$ and $M=80$ (four times more discretization steps).

In addition, the geometric path itself also plays an important role here. We have only explored in this work changing few geometric parameter, i.e. one or two bond lengths. The vast freedom to choose the geometric path indicates potential advantage. However, there needs further investigation on how to design an efficient and practical geometric path for more sophisticated cases. One possible approach is to define a coordinate of collective variables and identify the difficult cases that need the geometric path method as critical points along the coordinate. For instance, the reaction coordinate is often used to study chemical reactions, and the transition state is the maximum (or saddle) point on a one-dimensional (or two-dimensional) potential energy surface. When we need the geometric path method to reach the transition state, a path along the reaction coordinate could be a good choice.  But we leave these for future exploration.
\subsection{Effects of evolution time $T$ and discretization number $M$}
\begin{figure}[!h]
	\centering
	\includegraphics[width=0.47\textwidth]{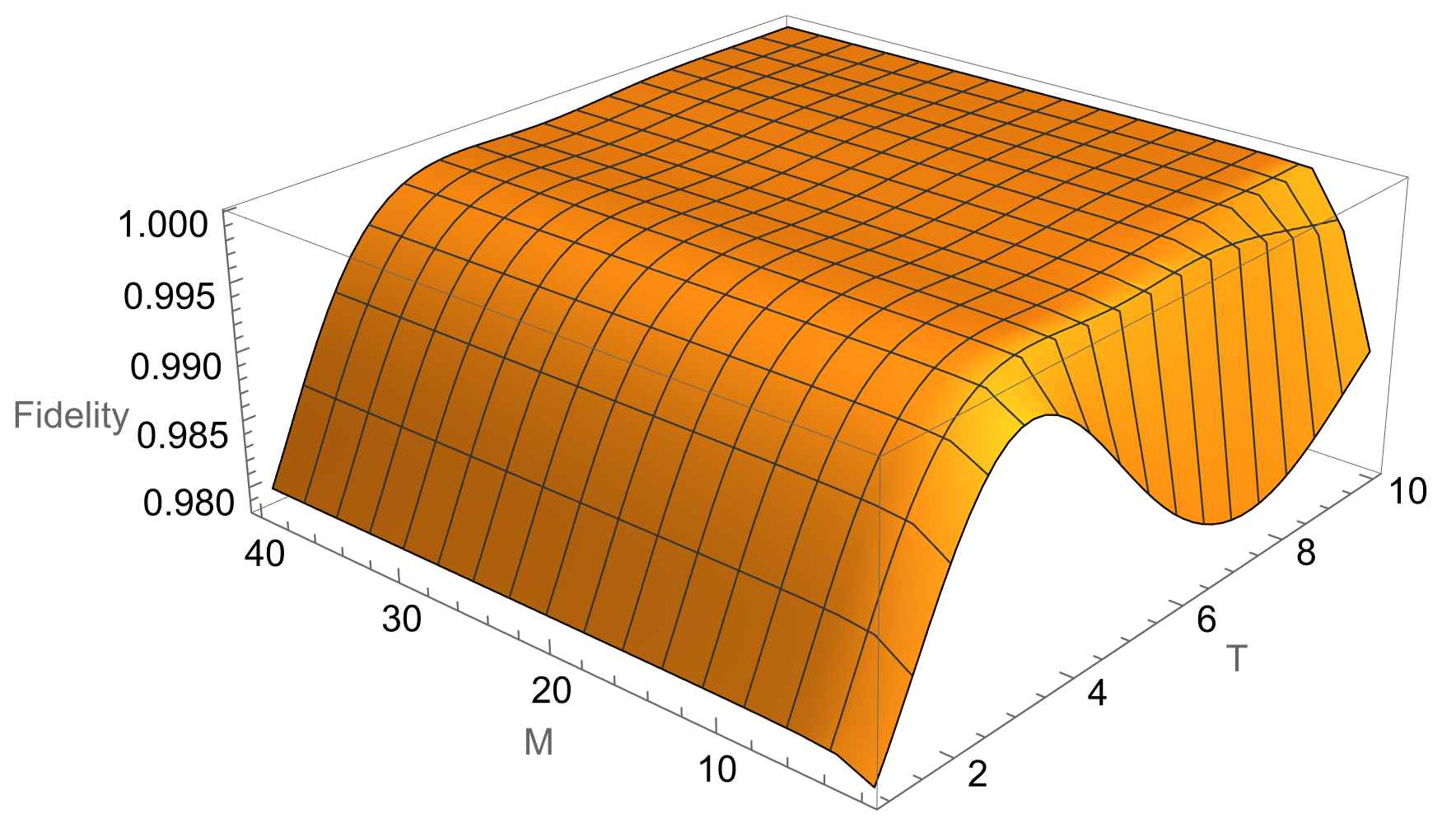}
	\caption{The results of the fidelity vs. $M$ and $T$ using the  H${}_2$O molecule as an illustration. The evolution is from $H_I^P$ to $H_{\text{full}}^P$ at O-H length $r=0.9584\AA$.}
	\label{h2ofTM}
\end{figure}
The two parameters $T$ and $M$ correspond to the the degree of adiabaticity and discretization approximating rate, respectively. If the adiabatic evolution does not fail (i.e. has a gap), we can theoretically achieve any accuracy by choosing sufficiently large $T$ and $M$. However, such large choices are not necessarily practical due to accumulation of noisy gate errors. In our simulations, we usually choose $T$ to be inverse proportional to the minimum gap of the evolution, and fixed the ratio of $T/M$ to keep the dicreatization error small. (However, one may not necessarily have the information on the gaps, and may need to empirically  test a few choices to see if the energy is lowered or converged as $T$ increases.) We tested the influence of $T$ and $M$ on the fidelity with the ground state and  the results are presented in the Fig.~\ref{h2ofTM}. In our study, we find that $T$ has larger impact than $M$ on the  fidelity.

\section{conclusion}\label{sec:conclude}
We have substantially improved an adiabatic algorithm for molecular energies (MC-QAE) introduced in a previous work by utilizing a geometric adiabatic evolution path (GeoQAE).  We showed by  numerical simulations of the adiabatic evolution that this new approach gives significant improvement on both ground-state energies and the fidelity of  wave functions (as well as lowest few excited states) for different molecules in the region where the MC-QAE fails due to the energy level crossing. We also demonstrated its potential application by simulating the potential surface for an exemplary chemical reaction involving two pairs of molecules. In our fidelity analysis, we find that our GeoQAE also gives high fidelity ground states using adiabatic evolution between two Hamiltonians  with finite difference in bond lengths.   The potential advantage of our method may rise from the vast freedom in choosing the adiabatic path.

The idea of choosing a suitable Hamiltonian that can be connected naturally and smoothly to the final Hamiltonian can be applied to other  many-body problems. In our present focus on the molecular energy, the MC Hamiltonian associated with a particular final molecular Hamiltonian generalizes the Fock operator such that the Hartree-Fock ground state  is an eigenstate with the eigenenergy being the Hartree-Fock ground-state energy.  

Although on the present noisy quantum devices, our GeoQAE approach might not yet be implemented to yield accurate results due to the required large gate numbers for the evolution operators, hardwares are continuing to be improved and this may make our algorithm practical in the future. Moreover, the existence of gaps in the suitably chosen geometric path may allow  alternative quantum approximate optimization algorithm (QAOA) like approaches or better variational ans\"atze to be developed.

\bigskip
\noindent
{\bf Acknowledgments}. 
Part of this work (in particular the research on the connection of the maximally commuting Hamiltonian to the Hartree-Fock solution) was supported  by the National Science Foundation  under Grant No. PHY 1915165.
The part of the research that involves quantum chemistry applications used resources of the 
Center for Functional Nanomaterials (CFN), which is a U.S. Department of Energy Office of Science User Facility, at the Brookhaven National Laboratory under Contract No. DE-SC0012704 by the U.S. Department of Energy,
Office of Science.
The research on the quantum algorithmic development in this work was partially supported by the  National Quantum Information Science
Research Centers under the 
``Co-design Center for Quantum Advantage''
award. 
\bibliography{bls}{}
\appendix
\section{mapping fermionic operators to qubits}
\label{appendix:mapping}
Here we briefly review two widely-used methods to map the fermionic operators to Pauli operators.
\subsection{Jordan-Wigner mapping}
The Jordan-Wigner transformation that we shall make from fermions ($a$'s) to 
spins ($\sigma$'s), 
\begin{eqnarray}
a_k =\big(\prod_{j=1}^k \sigma_j^z\big) (\sigma_k^x+ i \sigma_k^y)/2, \\
a_k^\dagger =\big(\prod_{j=1}^k \sigma_j^z\big) (\sigma_k^x- i \sigma_k^y)/2,
\end{eqnarray}
Let the spin state $\ket{-1}$ be qubit state $\ket{1}_q$ and $\ket{+1}$ be $\ket{0}_q$, it can be easily verified that the fermionic state $\ket{0}_f$ will be transformed to $\ket{0}_q$ and $\ket{1}_f$  to $\ket{1}_q$. Thus, the mapping keeps the wave function in the same expression. For example, the wavefunction in the fermionic basis $\ket{1011}_f$ will be mapped to qubit vector $\ket{1011}_q$.
\subsection{Parity mapping and qubit reduction}
Parity mapping is a method to map the wave function from the fermion-occupation basis to the so-called parity basis. Let $f_j$ (can only be 0 or 1 for fermions) be the occupation number in the fermionic basis, and $p_j=(\sum_{i=1}^j f_j \mod 2)$ counts the parity of the orbitals up to $j$, the parity basis can be chosen as $\ket{p_1 p_2...p_n}$. And the corresponding transformation is~\cite{seeley2012bravyi}
\begin{eqnarray}
a_k =\frac{1}{2}\left(\prod_{j=k+1}^n \sigma_j^x\right) (\sigma_k^x \sigma_{k-1}^z+ i \sigma_k^y), \\
a_k^\dagger =\frac{1}{2}\left(\prod_{j=k+1}^n \sigma_j^x\right) (\sigma_k^x \sigma_{k-1}^z - i \sigma_k^y).
\end{eqnarray}
Our label for fermions and qubits begin at 1 and we define the operator  $\sigma_{0}^z$ to be the identity matrix. One can verify that the fermionic basis $\ket{f_1...f_n}$ can be mapped to the parity basis $\ket{p_1...p_n}$ according to the definition of $p_i$s. For example, the fermionic state $\ket{101110}_f$ will be mapped to parity state $\ket{110100}_p$.

Now consider a fermionic state with spin configuration $\ket{f_{1}^{\alpha}...f_{n}^{\alpha},f_{n+1}^{\beta}...f_{2n}^{\beta}}$ and its parity correspondence $\ket{p_{1}...p_{n},p_{n+1}...p_{2n}}$. Note that the $p_n$ counts for the parity of all $\alpha$-spin orbitals and $p_{2n}$ counts for the parity of all orbitals. Therefore, if one further assume the conservation of the spin and particle number, the parity numbers $p_n,~ p_{2n}$ are constants. Thus, one can remove the two qubits and only consider the states in the subspace spanned by states with same electron number and spin configuration, which reduces the complexity of both quantum computations and classical simulations.
\section{Brief review of Hartree-Fock method}
\label{app:HF}
Here we summarize the Hartree-Fock procedure so as to introduce  the notation used in the main text of the paper. We refer the readers to standard textbooks such as, Ref.~\cite{szabo2012modern}, for more details. Given a molecule, our goal is to solve the time-independent electronic Schr\"odinger equation,
\begin{widetext}
\begin{equation}
\left[-\frac{1}{2} \sum_{i} \nabla_{i}^{2}-\frac{1}{2} \sum_{I} \frac{\nabla_{I}^{2}}{M_I/m_e}-\sum_{A, i} \frac{Z_{A}}{r_{A i}}+\sum_{A>B} \frac{Z_{A} Z_{B}}{R_{A B}}+\sum_{i>j} \frac{1}{r_{i j}}\right] \Psi(\mathbf{r} ; \mathbf{R})=E \Psi(\mathbf{r} ; \mathbf{R}),
\end{equation}
\end{widetext}
where $Z_{I}$ is the atomic number for atom I (whose nucleus shares the same symbol and has mass $m_I$), $i$ and $j$ index electrons (whose mass is denoted by $m_e$ and charge $e$ is set to unity).  If we are only interested in the electronic wave functions, we can treat the nuclei as fixed point charges, and the Hamiltonian will be simplified to
\begin{widetext}
\begin{equation}
\left[-\frac{1}{2} \sum_{i} \nabla_{i}^{2}-\sum_{A, i} \frac{Z_{A}}{r_{A i}}+\sum_{i>j} \frac{1}{r_{i j}}\right] \Psi(\mathbf{r} ; \mathbf{R})=E_{el} \Psi(\mathbf{r} ; \mathbf{R}).
\end{equation}
\end{widetext}
Suppose we use single Slater-determinants as one-electron wave function, and introduce fermionic creation operators and annihilation operators, the Hamiltonian can be written in the second quantized picture
\begin{equation}
H_{el}=\sum_{i, j} t_{ij} a_{i}^{\dagger} a_{j}+\frac{1}{2} \sum_{i,j,k,l} u_{ijkl} a_{i}^{\dagger} a_{k}^{\dagger} a_{l} a_{j}
\end{equation}
where
\begin{equation}
t_{ij}=\langle i|h| j\rangle=\int d \mathbf{x}_{1} \Psi_{i}^{*}\left(\mathbf{x}_{1}\right) \left(-\frac{1}{2} \nabla_{1}^{2}-\sum_{A} \frac{Z_{A}}{r_{A 1}}\right) \Psi_{j}\left(\mathbf{x}_{1}\right)
\end{equation}
and
\begin{eqnarray}
u_{ijkl}&=&[i j \mid k l]\\
&=&\int d \mathbf{x}_{1} d \mathbf{x}_{2} \Psi_{i}^{*}\left(\mathbf{x}_{1}\right) \Psi_{j}\left(\mathbf{x}_{1}\right) \frac{1}{r_{12}} \Psi_{k}^{*}\left(\mathbf{x}_{2}\right) \Psi_{l}\left(\mathbf{x}_{2}\right)\nonumber
\end{eqnarray}
were introduced in the main text. The energy expression becomes
\begin{eqnarray}
\nonumber
E_{e l}&=&\left\langle\Psi\left|\hat{H}_{e l}\right| \Psi\right\rangle\\
&=&\sum_{i}\langle i|h| i\rangle+\frac{1}{2} \sum_{i j}([i i \mid j j]-[i j \mid j i])
\end{eqnarray}
In Hartree-Fock method, we assume the ground state wave function can be approximated by a single Slater determinant, and the Hamiltonian can be reduced to a one-body Hamiltonian,
\begin{eqnarray}
\label{eq:Fock}
\hat{f}=\sum_{i, j}\left(t_{i j}+V_{i j}\right) a_{i}^{\dagger} a_{j}, \\
V_{i j}=\sum_{k \in o c c}\left(u_{i k k j}-u_{i k j k}\right),
\end{eqnarray}
where $\hat{f}$ is referred to as the Fock operator. 
In the Hartee-Fock calculation, we seek for an optimal molecular basis set (labeled by Greek letters in the main text) that can minimize the energy expectation value, starting from a chosen set of atomic orbitals (labelled by Roman letters), such as the STO-3G basis set. An iterative calculation can be performed to obtain the optimal molecular basis set. After we obtain one basis set for the Fock operator, we diagonalize the operator and rewrite the full Hamiltonian in the new basis set, then perform the diagonalization again until the result converges.

We use the quantum chemistry package {\tt PySCF}~\cite{PYSCF} to perform this classical step of calculations.

\section{Jordan-Wigner mapping of the initial Hamiltonian and its relation to the Hartree-Fock solution}
\label{appendix:HF}
We use Jordan-Wigner mapping as an example to show the connection between our choice of the initial Hamiltonian and the classical Hartree-Fock method. Suppose after a classical Hartree-Fock calculation, we get a set of converged orbital basis $\{\ket{\psi_{\alpha}}\}$ and their corresponding creation and annihilation operators $a_{\alpha},~a_{\alpha}^{\dagger}$, the full Hamiltonian in the second quantization picture is
\begin{eqnarray}
	H_{\text{full}}^F&=&\sum_{\alpha,\beta} t_{\alpha \beta}^\text{MO} a_{\alpha}^{\dagger} a_{\beta}+\frac{1}{2} \sum_{\alpha, \beta,\gamma, \delta} u_{\alpha \beta \gamma \delta}^\text{MO} a_{\alpha}^{\dagger} a_{\gamma}^{\dagger} a_{\delta} a_{\beta},\nonumber
\end{eqnarray}
which, by Jordan-Wigner mapping, results in a summation of Pauli operators $H_P=\sum P_i$. And our choice of the initial Hamiltonian for the Pauli operators is the summation only over the terms containing only $Z$s and $I$s, which originate from the terms of the fermionic operators that only contains number operators $a^{\dagger}a$,
\begin{eqnarray}
\label{eq:HIa}
	H_I^{F}&=&\sum_{\alpha} t_{\alpha \alpha}^\text{MO} a_{\alpha}^{\dagger} a_{\alpha}+
	\\
	&&\frac{1}{2} \sum_{\alpha, \beta} (u_{\alpha \alpha \beta \beta}^\text{MO} a_{\alpha}^{\dagger} a_{\beta}^{\dagger} a_{\beta} a_{\alpha}-u_{\alpha \beta \beta \alpha}^\text{MO}a_{\alpha}^{\dagger} a_{\beta}^{\dagger} a_{\alpha} a_{\beta}),\nonumber
\end{eqnarray}
Note that due to the choice of our molecular basis, the $\ket{\psi_{\alpha}}$s are converged Hartree-Fock molecular basis. Thus, the Hartree-Fock state in the fermionic basis is simply occupying the lowest energy orbitals. Suppose the wavefunction for orbital labels $f_1^\alpha, f_2^\alpha,...$ with lowest energy first (for same spins) and consisting of $\alpha$ and $\beta$ spin takes the form $\ket{f_{1}^\alpha...f_{n}^\alpha,f_{n+1}^{\beta}...f_{2n}^{\beta}}$, the Hartree-Fock wavefunction for a four-electron and eight-orbital state is 
$\ket{1100,1100}$. It is obvious that the state is one of the eigenstate of our initial Hamiltonian \eqref{eq:HIa}, with corresponding eigenenergy
\begin{eqnarray}
&~&H_I^{F}\ket{\psi_{HF}}\\
&=&\left(\sum_{\alpha\in occ}t_{\alpha\alpha}^\text{MO}+\frac{1}{2} \sum_{\alpha, \beta\in occ}(u_{\alpha \alpha \beta \beta}^\text{MO}-u_{\alpha \beta \beta \alpha}^\text{MO})\right)\ket{\psi_{HF}},\nonumber
\end{eqnarray}
Comparing the value with the expression of the Hartree-Fock energy \eqref{eq:Ehf}, one can verify that they are exactly the same. Therefore, the Hartree-Fock wave function is an eigenstate of our initial Hamiltonian with the eigenenergy equal to the Hartree-Fock energy. We remark that the Hartree-Fock state is not always the ground state of the $H_I^F$ (while in most cases where the bond lengths are near equilibrium, it is), and the fermionic Hamiltonian $H_I^F$ is also different from the conventional definition of Hartree-Fock Hamiltonian \eqref{eq:Fock}, where only one-body terms are contained.
\section{Methods to ensure the continuity of the Hamiltonian}
\label{continuity}
The energy gap for the evolution from $H_{\text{full}}^P(\bm{r})$ to $H_{\text{full}}^P(\bm{r}+\Delta \bm{r})$ rely on the continuity of the Hamiltonian with respect to geometric configuration $\bm{r}$. If the Hamiltonian difference $|H_{\text{full}}^P(\bm{r})-H_{\text{full}}^P(\bm{r}+\Delta \bm{r})|$ is small, then the changes of the spectrum are also small, which gives rise to a relatively large energy gap during the evolution so that the evolving state is not likely to mix with other energy levels; see e.g. Fig.~\ref{H2Ogap}b. Since the $H_{\text{full}}^P(\bm{r})$ is transformed from the fermionic Hamiltonian $H_{\text{full}}^F(\bm{r})$, the continuity of $H_{\text{full}}^P(\bm{r})$ is equivalent to that of $H_{\text{full}}^F(\bm{r})$, which is essentially the continuity of $t_{\alpha \beta}^\text{MO} (\bm{r})$ and $u_{\alpha \beta \gamma \delta}^\text{MO} (\bm{r})$. If we choose a fixed basis for all configurations of a molecule, the continuity is guaranteed by the physical continuity of the Hamiltonian with respect to the geometric configuration $\bm{r}$. However, our choice for each configuration is the Hartree-Fock molecular basis, which depends on the Hartree-Fock calculation at each geometric configuration and is not necessarily continuous with bond length. We will discuss below several methods to ensure the continuity of the $H_{\text{full}}^P(\bm{r})$. We emphasize again that for our purpose it is not important how good the  Hartree-Fock solution is to the exact solution but continuity  of the Hamiltonians $H^P_\text{full}(\bm{r})$ as $\bm{r}$ varies.
\subsection{Initial guess}
The efficiency and convergence of a Hartree-Fock calculation depends largely on the choice of the initial guess. In our algorithm, we need the Hartree-Fock molecular bases for a series geometric configurations $\bm{r}_0,\bm{r}_1,...,\bm{r}_N$, where $\bm{r}_0$ is chosen to be a near-equilibrium configuration and the distances between each successive configurations $|\bm{r}_{n+1}-\bm{r}_n|$ are small. Since the difference of the two configurations is small, the Hartree-Fock calculation results of them are close to each other. Thus, the converged basis from previous configuration is a good initial guess for the next configuration and a Hartree-Fock calculation is likely to converge near the initial basis. 

Therefore, for a given configuration $\bm{r}_n$ of the molecule, our choice of initial guess is chosen to be the converged density matrix at previous configuration $\bm{r}_{n-1}$. For the initial configuration $\bm{r}_0$, we can choose any initial guess that gives a converged result. 
\subsection{Permutations and signs of the molecular basis}
Suppose we have obtained  a set Hartree-Fock molecular basis states $\{\ket{\varphi_i}_{\bm{r}}\}$ at configuration $\bm{r}$, it is obvious that, if we perform a permutation $\ket{\varphi_1'}=\ket{\varphi_2},~\ket{\varphi_2'}=\ket{\varphi_1}$, or add a minus sign $\ket{\varphi_1'}=-\ket{\varphi_1}$, the new molecular basis set is equivalent as before. However, after one does such transformation to the basis, the Hamiltonian should also be transformed accordingly. And such transformation will result in a noticeable non-vanishing increase of the difference of the Hamiltonians $|H_{\text{full}}^P(\bm{r})-H_{\text{full}}^P(\bm{r}+\Delta r)|$ (compared to the case where the bases at both distances are very similar, even in terms of their ordering) despite that  $\Delta \bm{r}$ is very close to 0.

To solve this problem, we manually perform a proper permutation and a suitable sign change (if necessary) to re-align and enforce the molecular basis $\{\ket{\varphi_i}_{\bm{r}+\Delta \bm{r}}\}$ at $\bm{r}+\Delta r$ consistent with $\{\ket{\varphi_i}_{\bm{r}}\}$ at $\bm{r}$, so that the basis  is continuous with respect to the change in the geometric configuration $\bm{r}$. In our calculation, the molecular orbitals $\ket{\varphi_i}$ are written in combinations of atomic orbitals $\ket{\varphi_i}=\sum_j C_{ij} \ket{\phi_j^{\text{AO}}}$, where the superscript AO denotes the atomic orbitals, which are constructed locally and automatically continuous with geometric configuration $\bm{r}$. Therefore, by comparing and re-aligning the molecular coefficients $C_{ij}^{(\bm{r})}$ and $C_{ij}^{(\bm{r}+\Delta \bm{r})}$ at the successive configurations $\bm{r}, ~\bm{r}+\Delta\bm{r}$, one can ensure the continuity of the Hartree-Fock molecular basis with respect to the geometric configuration $\bm{r}$.
\subsection{Transforming the unconverged molecular basis}
In some cases where the Hartree-Fock calculation is hard to converge and cannot be improved by choosing a good initial guess, we need to abandon the quality of the Hartree-Fock results to ensure the continuity of the Hamiltonian $H^P_\text{full}(\bm{r})$. To do this, we use the molecular coefficients $C_{ij}^{(\bm{r}_n)}$ at the previous configuration $\bm{r}_{n}$ to calculate the molecular basis at next configuration $\bm{r}_{n+1}$
\begin{eqnarray}
    \ket{\varphi_i}_{\bm{r}_{n+1}}\simeq\sum_j C_{ij}^{(\bm{r}_{n})}\ket{\phi_j^{\text{AO}}}_{\bm{r}_{n+1}}.
\end{eqnarray}
Note that the atomic orbitals are not necessarily orthogonal to each other. We need additional transformations to ensure $\ket{\varphi_i}_{\bm{r}_{n+1}}$ is an orthonormal basis. Denote $S_{ij}^{(\bm{r}_n)}=\braket{\phi_i^{\text{AO}}|\phi_j^{\text{AO}}}_{\bm{r}_n}$ and $S_{ij}^{(\bm{r}_{n+1})}=\braket{\phi_i^{\text{AO}}|\phi_j^{\text{AO}}}_{\bm{r}_{n+1}}$, and assume the molecular basis at $\bm{r}_n$ is already orthonormal $\braket{\varphi_i|\varphi_j}_{\bm{r}_n}=\delta_{ij}$. We construct the molecular basis at $\bm{r}_{n+1}$:
\begin{eqnarray}
\nonumber
    \ket{\varphi_i}_{\bm{r}_{n+1}}&=&\sum_{j,k,l} C_{ij}^{(\bm{r}_{n})}(S^{(\bm{r}_{n})})_{jk}^{\frac{1}{2}}(S^{(\bm{r}_{n+1})})_{kl}^{-\frac{1}{2}}\ket{\phi_l^{\text{AO}}}_{\bm{r}_{n+1}}\\
    \label{eq:newMO}
    &\equiv&\sum_{l} D_{il}^{(\bm{r}_{n+1})}\ket{\phi_l^{\text{AO}}}_{\bm{r}_{n+1}}.
\end{eqnarray}
where the matrix power operation $A^{1/2}$ is defined as follows: if matrix $A$ can be diagonalized $A=U^{T}\Lambda U$ and $\Lambda$ is a positive diagonal matrix, we define $A^{1/2}=U^T \Lambda^{1/2} U$. One can easily verify that $\braket{\varphi_i|\varphi_j}_{\bm{r}_{n+1}}=\delta_{ij}$. Note that matrix $S^{(\bm{r})}$ is only related to $\ket{\phi^{\text{AO}}}_{\bm{r}}$. If $\Delta \bm{r}=\bm{r}_{n+1}-\bm{r}_n$ is small enough, we have $\ket{\phi_l^{\text{AO}}}_{\bm{r}_{n+1}}\sim \ket{\phi_l^{\text{AO}}}_{\bm{r}_{n}}$ and $S^{(\bm{r}_n)}\sim S^{(\bm{r}_{n+1})}$. Thus,
\begin{equation}
    \ket{\varphi_i}_{\bm{r}_{n+1}}\sim\sum_j C_{ij}^{(\bm{r}_{n})} \ket{\phi_l^{\text{AO}}}_{\bm{r}_{n}}=\ket{\varphi_i}_{\bm{r}_{n}}.
\end{equation}
 After transforming the Hamiltonian $H^P_\text{full}(\bm{r}_{n+1})$ according to the new molecular basis \eqref{eq:newMO}, we assure the continuity of the Hamiltonian.

\end{document}